\documentclass[%
 aip,
 amsmath,amssymb,
preprint,%
]{revtex4-1}
\usepackage{graphicx}
\usepackage{dcolumn}
\usepackage{bm}
\usepackage{color}

\usepackage[utf8]{inputenc}
\usepackage[T1]{fontenc}
\usepackage{mathptmx}
\usepackage{subfig}
\usepackage[inline]{enumitem}
\usepackage{caption}
\usepackage{float}
\newcommand{\ppx}[2]{\frac{\partial#1}{\partial#2}}

\renewcommand{\vec}{\mathbf}

\usepackage[normalem]{ulem}

\begin{document}


\title[]{Resolving the mystery of electron perpendicular temperature spike in the plasma sheath}
\author{Yanzeng Zhang}%
\affiliation{Theoretical Division, Los Alamos National Laboratory, Los Alamos, New Mexico 87545, USA}
\author{Yuzhi Li}
\affiliation{Kevin T. Crofton Department of Aerospace and Ocean Engineering, Virginia Tech, Blacksburg, Virginia 24060, USA}
\author{Bhuvana Srinivasan}
\affiliation{Kevin T. Crofton Department of Aerospace and Ocean Engineering, Virginia Tech, Blacksburg, Virginia 24060, USA}
\author{Xian-Zhu Tang}%
\affiliation{Theoretical Division, Los Alamos National Laboratory, Los Alamos, New Mexico 87545, USA}


\begin{abstract}

A large family of plasmas has collisional mean-free-path much longer
than the non-neutral sheath width, which scales with the plasma Debye
length. The plasmas, particularly the electrons, assume strong
temperature anisotropy in the sheath. The temperature in the sheath
flow direction ($T_{e\parallel}$) is lower and drops towards the wall
as a result of the decompressional cooling by the accelerating sheath
flow.  The electron temperature in the transverse direction of the
flow field ($T_{e\perp}$) not only is higher but also spikes up in the
sheath.  This abnormal behavior of $T_{e\perp}$ spike is found to be
the result of a negative gradient of the parallel heat flux of
transverse degrees of freedom ($q_{es}$) in the sheath. The non-zero
heat flux $q_{es}$ is induced by pitch-angle scattering of electrons
via either their interaction with self-excited electromagnetic waves
in a nearly collisionless plasma or Coulomb collision in a
collisional plasma, or both in the intermediate regime of plasma
collisionality.

\end{abstract}

\maketitle

\section{Introduction}

When a plasma is in contact with solid boundaries, due to the greater
mobility of electrons, a non-neutral plasma sheath forms next to the
wall\cite{Langmuir,langmuir-pra-1929,lieberman-lichtenberg-book-2005,stangeby-book-2000}.
In the absence of the copious amount of electron emission from the
wall, a negative electrical potential is established at the boundary,
which is promptly shielded out over a few Debye lengths $\lambda_D$ so
that the bulk plasma remains quasi-neutral.  The resulting sheath
electric field is essential for maintaining ambipolar transport,
through which the particle and heat losses from the plasma to the
solid boundary are regulated.

In most low-density plasmas of interest, the collisional
mean-free-path ($\lambda_{mfp}$) is larger or much larger than the
Debye length, so the sheath Knudsen number $Kn^{sh}\equiv
\lambda_{mfp}/\lambda_D,$ which is defined as the ratio between the
plasma mean free path and the Debye length at the sheath entrance,
satisfies $Kn^{sh}>1$ or $Kn^{sh}\gg 1.$ Remarkably, despite
$Kn^{sh}\gg 1,$ large gradients of plasma temperature, density, and
flow can be sustained in the narrow sheath region on the order of a
few Debye lengths. This is fundamentally the result of the large
sheath electric field, itself a large gradient of the sheath
electrostatic potential, which is independent of the collisional
mean-free-path, which would otherwise set the gradient length scale in
the quasi-neutral plasma away from the sheath region.

The lack of plasma collisions in the narrow sheath region allows
strong temperature anisotropy to develop. The driver is the sheath
plasma flow into the wall, which has a large gradient along the
streamline direction due to the sheath electric field acceleration of
the mostly collisionless ions.  Let's label the wall-bound plasma flow
direction as parallel and the cross-flow plane as perpendicular, and
define two temperatures $T_\parallel$ and $T_\perp.$ The accelerating
sheath flow would decompressionally cool $T_\parallel,$ so a
temperature anisotropy of $T_\parallel < T_\perp$ would naturally
develop in the sheath region.~\cite{tang2011kinetic} In a strongly
magnetized plasma where the magnetic field intercepts the wall at a
large angle, the plasma mostly flows along the magnetic field line.
This translates into a much lower parallel electron temperature
$T_{e\parallel}$ compared with the perpendicular electron temperature
$T_{e\perp},$ all with respect to the magnetic field $\mathbf{B}.$ The
situation becomes more complicated when the magnetic field line
intercepts the wall at an oblique angle,~\cite{chodura-sheath} for
which the plasma flow in the Debye sheath would be non-aligned with
the magnetic field as it meets the wall. In that case, as
$T_{\parallel,\perp}$ originally defined with respect to the flow
direction, they become non-aligned with the magnetic field as well.

\textcolor{red}{It can be noted that the interesting physics of anisotropic
temperature for the sheath plasma is usually ignored in the vast
plasma sheath literature (for recent reviews, see
Refs.~\onlinecite{robertson-ppcf-2013,riemann-psst-2008,franklin-JPDAP-2003}).
This is simply the result of deploying physical models with isotropic
plasma temperatures.  First principle kinetic simulations of plasma
sheath, using either particle-in-cell (e.g.,
Ref.~\onlinecite{tang2011kinetic}) or continuum discretization (e.g.,
Ref.~\onlinecite{skolar2022continuum}) would be able to capture the
temperature anisotropy physics, but only if the perpendicular degrees
of freedom in the momentum space are accounted for. For strictly 1D1V
kinetic modeling, such as that reported in
Ref.~\onlinecite{sun-etal-FiP-2022}, the physics of sheath plasma
temperature anisotropy would still be excluded.}

While the deep drop of $T_{e\parallel}$ in the sheath region is well
understood as the result of decompressional cooling by the
accelerating sheath flow~\cite{tang2011kinetic,tang_kinetic2016}, there is a
long-standing mystery in the behavior of $T_{e\perp}$ in the plasma
sheath. Instead of staying flat or slowly dropping, $T_{e\perp}$
spikes up in the non-neutral plasma sheath in an unmagnetized plasma
or a magnetized plasma with a large angle between the magnetic field
and the wall.  This is most clearly demonstrated in first-principles
kinetic
simulations~\cite{tang-guo-pop-2015,li2022bohm,li2022transport}, as
all physical quantities are readily diagnosed from simulation data. We
shall note that although the evidence to such effect was explicitly
reported in Ref.~\onlinecite{tang-guo-pop-2015} (see Figs.~1-2 in the
paper) as an unresolved mystery and more recently in
Refs.~\onlinecite{li2022bohm,li2022transport} (see Fig.~1 in the
Supplemental Material of Ref.~\onlinecite{li2022bohm} and Fig.~2 in
Ref.~\onlinecite{li2022transport}), all in the context of VPIC
simulations, there is a good reason to believe that others must have
encountered the same mystery in the first-principles kinetic
simulations where the physics of $T_{e\perp}$ degrees of freedom are
retained \textcolor{red}{(plausible examples include Fig.~9 in
  Ref.~\onlinecite{skolar2022continuum} where kinetic simulations with
  continuum discretization as opposed to particle-in-cell method is
  deployed, and Fig. 13 in Ref.~\onlinecite{janhunen2018evolution}
  where electrostatic 2D particle-in-cell simulations was performed
  for $\mathbf{E}\times{\mathbf{B}}$ plasmas)}.

This paper aims to elucidate the underlying physics that would resolve
the mystery of $T_{e\perp}$ spike in the plasma sheath by considering
a normal $\mathbf{B}$ to the walls. We will show that such behavior is
associated with a \textit{negative gradient of the parallel electron
  heat flux of the perpendicular degrees of freedom, $q_{es}$}. As
defined in the original formalism by Chew \textit{et al}~\cite{Chew1956},
\begin{align}
q_{es} \equiv (1/2) \int m_e w_\perp^2 w_\parallel f_e
d^3\mathbf{v} \label{eq:qes-def}
\end{align}
with $\mathbf{v}$ the particle velocity, $\mathbf{w}$ the electron
peculiar velocity, $w_\parallel \equiv \mathbf{w}\cdot{\mathbf{b}},$
$w_\perp\equiv \mathbf{w} - w_\parallel\mathbf{b},$ and $\mathbf{b}
\equiv \mathbf{B}/B.$ This can be compared with the parallel electron
heat flux of parallel degrees of freedom, which takes the form
\begin{align}
q_{en} \equiv \int m_e w_\parallel^2 w_\parallel f_e d^3\mathbf{v}.
\end{align}
Parallel streaming loss in the neighborhood of the plasma sheath leads
to a truncation of the electron distribution function in the direction
that is opposite to the sheath flow, the asymmetry of which yields a
wall-bound $q_{en}$ even in the absence of collisions and
wave-particle interaction. In sharp contrast, finite pitch-angle
scattering is required to both isotropize the parallel and
perpendicular electron temperature, and produce a finite $q_{es}.$
There are two mechanisms for the pitch-angle scattering of electrons in
the neighborhood of the plasma sheath: one is Coulomb collisions,
which is the reason for $T_{e\perp}$ spike in the sheath of
collisional plasma that has $Kn^{sh} > 1$ but not $Kn^{sh}\gg
1$~\cite{tang-guo-pop-2015,li2022bohm,li2022transport}; and the other
is resonant wave-electron interaction in a nearly collisionless plasma
with $Kn^{sh}\gg 1.$ The self-excited wave instability could be
whistler waves in a magnetized
plasma~\cite{kennel1966limit,guo2012ambipolar} or Weibel instability
in an unmagnetized plasma~\cite{tang2011kinetic}. It should be noted
that, although the effects of whistler waves/turbulence on the
electron particle flux~\cite{kennel1966limit,guo2012ambipolar} and
heat flux of the parallel degree of
freedom~\cite{roberg2018suppression}, $q_{en}$, have been well
documented, its role in $q_{es}$ modulation and thus a spike of
$T_{e\perp}$ in the sheath of a nearly collisionless plasma have not
been reported yet.

The subtle physics of $T_{e\perp}$ spike in the sheath region can be
demonstrated in the archetypal example of a one-dimensional (1D)
three-velocity (3V) plasma in a slab geometry with a strong magnetic
field normal to the absorbing boundaries. To compensate for the
particle loss to the walls, we introduce an upstream source that draws
from a local Maxwellian of fixed source temperature. As a result, a
steady state can be sustained. We must emphasize that the rise of
$T_{e\perp}$ towards the wall appears not only in the steady state but
also in its early time evolution process. However, the steady state
will be employed to illustrate the role of $q_{es}$ in the
$T_{e\perp}$ spike. Moreover, it is worth noting that the
aforementioned two mechanisms for the pitch-angle scattering of
electrons have different electrostatic/electromagnetic natures in that
the whistler waves can be excited only in electromagnetic simulations
while the collisions work equally in the electromagnetic and
electrostatic simulations. Indeed, the first-principles kinetic
simulations using VPIC~\cite{VPIC} confirm no differences, even in the
sheath region, of strongly collisional plasmas between the
electromagnetic and electrostatic simulations. Before elucidating the
underlying physics for the $T_{e\perp}$ spike in the sheath, we
briefly discuss the VPIC simulation setup, which is similar to that in
Refs.~\onlinecite{li2022bohm,li2022transport}.
\textcolor{red}{Specifically, a uniform proton-electron plasma with
  density $n_0$ and temperature $T_0$ is initially filled in the
  simulation domain, and a strong magnetic field is introduced so that
  the plasma has a low-$\beta$ with $\beta\approx 1.4$\%. The plasma
  source with temperature $T_0$ is in the middle of the domain $x\in
  [3L/8, 5L/8]$ with two absorbing walls at $x=0$ and $x=L\equiv
  256\lambda_D$.  The resolution of the simulation is $\Delta x
  =0.4\lambda_D$ with 2500 macro-particles per cell (note that same
  results are obtained for simulations with higher resolution, $\Delta
  x =0.1\lambda_D$, and number of macro-particles, 10000). For
  collisional plasmas, we will use an artificial Coulomb logarithm
  $\ln \Lambda$ to obtain different collisional regimes characterized
  by the nominal Knudsen number $Kn=\lambda_{mfp}/\lambda_D$, where
  Takizuka and Abe's method~\cite{T.A} is employed as the collisional
  model in VPIC. }
  
The rest of the paper is organized as follows. Sections \ref{sec-coll}
and \ref{sec-non-coll} will consider, respectively, a collisional and
collisionless plasma with a normal $\mathbf{B}$ to the walls, where
$T_{e\perp}$ spikes near the wall. In section \ref{sec-oblique-B} we
will discuss a collisional plasma in an oblique magnetic field with a
small angle to the walls, which would provide additional decompressional cooling to $T_{e\perp}$ (with respect to the plasma flow direction) so
that $T_{e\perp}$ spike would
disappear. Section \ref{sec-conclude} will conclude.

\section{Collisional plasma with a normal magnetic field to the walls \label{sec-coll}}

\begin{figure}[tbh]
\centering
\includegraphics[width=0.6\textwidth]{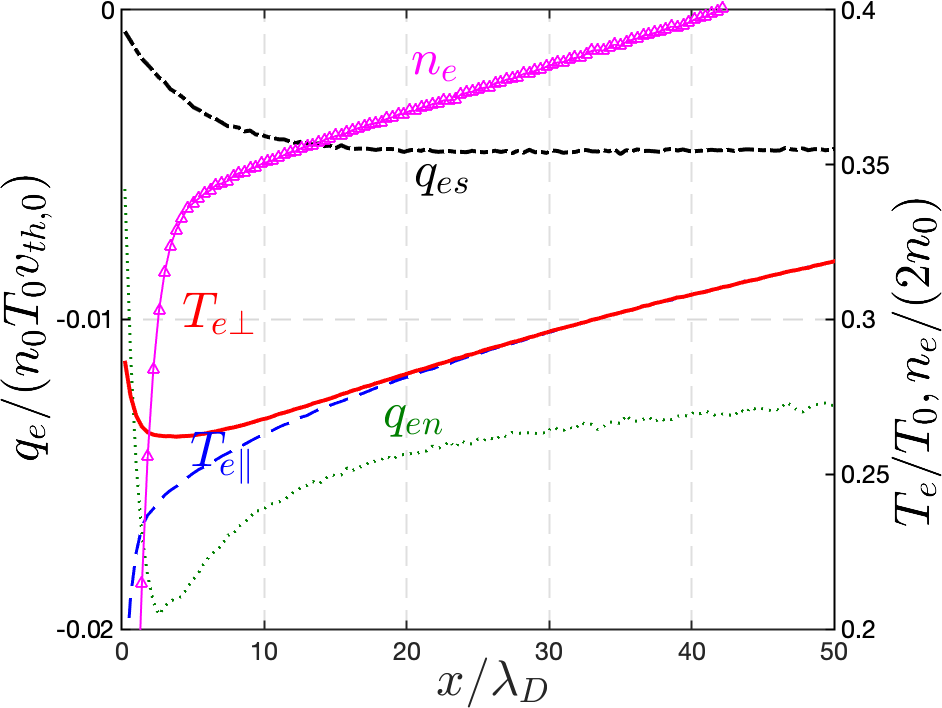}
\caption{The electron temperature \textcolor{red}{and density}
  (corresponding to the right y-axis) and heat flux (corresponding to
  the left y-axis) in a steady state from the first-principles
  simulation using VPIC~\cite{VPIC}. The nominal Knudsen number $Kn$,
  defined as the ratio of the initial electron mean free path to the
  Debye length, is $Kn=20$. $n_0,~T_{0}$ and $v_{th,0}$ are initial
  plasma density, temperature and electron thermal speed. The
  time-averaging (but not spatial-averaging) employed in
  Ref.~\onlinecite{li2022bohm} over a long period in the steady state
  is utilized to overcome the PIC noise. We note that the
  electrostatic and electromagnetic simulations provide the same
  results so only the former are plotted here.}
\label{fig:Teperp-qes-kn-20}
\end{figure}

In a strongly collisional plasma (but $Kn^{sh}>1$ is still satisfied),
the kinetic instabilities like the whistler instability are
suppressed~\cite{zhangcolli}. As a result, the magnetic field is
unperturbed such that the parallel direction is still along $x$ and
$\nabla \cdot \hat{b} = 0$. This indicates that the $T_{e\perp}$ spike
in the sheath has an electrostatic characteristic in a collisional plasma (e.g., see
Fig.~\ref{fig:Teperp-qes-kn-20}). As a result,
the anisotropic energy equations for electrons in the steady state
sheath region read~\cite{Chew1956, Chodura_1971}
\begin{equation}     
  	n_eu_{ex}\ppx{T_{e\parallel}}{x}+2n_eT_{e\parallel}\ppx{u_{ex}}{x}+\ppx{q_{en}}{x}
        = Q_{ee}+Q_{ei}, \label{eq:transport-equation-Tex}
\end{equation}
\begin{equation}     
  	n_eu_{ex}\ppx{T_{e\perp}}{x}+\ppx{q_{es}}{x}  = -(Q_{ee}+Q_{ei})/2, \label{eq:transport-equation-Tey}
\end{equation}
where 
$n_e$ is the electron density and $u_{ex}$ is the parallel electron
flow. Notice that we use $\parallel$ in the subscript only for the
parallel electron temperature (to distinguish from $T_{e\perp}$) while
using $x$ for the parallel direction in all the other quantities. For
the energy transfer due to collisions, we only keep the dominant
temperature isotropization terms $Q_{ee,ei}$ for illustration
purpose~\cite{li2022bohm}, where $Q_{ee,ei}$ denote electrons
colliding with electrons and ions, respectively.

In the regime of $Kn^{sh}>1,$ the collisions in the sheath region are
sufficiently weak that $Q_{ee,ei}$ are subdominant in
Eqs.~(\ref{eq:transport-equation-Tex}, \ref{eq:transport-equation-Tey}).
Ignoring them, one finds the remaining difference between the two
equations is the presence of the decompressional cooling term $\partial u_{ex}/\partial x$ in
Eq.~(\ref{eq:transport-equation-Tex}), which is absent in
Eq.~(\ref{eq:transport-equation-Tey}).  For $T_{e\parallel},$ the
decompressional cooling term overwhelms the conduction flux
contribution, so
\begin{align}
n_e u_{ex} \frac{\partial T_{e\parallel}}{\partial x} \approx - 2 n_e
T_{e\parallel}\frac{\partial u_{ex}}{\partial x},
\end{align}
which says that decompressional cooling due to an accelerating sheath
flow produces a decreasing $T_{e\parallel}$ as the plasma approaches
the wall.~\cite{tang_kinetic2016} It is important to note that this
happens despite the wall-bound $q_{en}$ heat flux drops in magnitude
towards the wall, which contributes a heating mechanism for
$T_{e\parallel},$ except that it is simply too weak compared with
decompressional cooling.  For $T_{e\perp},$ the gradient of heat flux
$q_{es}$ is the only term that drives $T_{e\perp}$ variation,
\begin{align}
n_e u_{ex} \frac{\partial T_{e\perp}}{\partial x} \approx -
\frac{\partial q_{es}}{\partial x}.\label{eq:Tey-qes-coll}
\end{align}
Because of this, as the non-neutral sheath reduces the wall-bound heat
flux $q_{es},$ like what it does to $q_{en},$ the heating
effect that $\partial q_{es}/\partial x$ brings, would heat up the
$T_{e\perp}$ in the sheath.

In the mostly collisionless sheath of a collisional bulk plasma, we
have previously shown~\cite{tang_kinetic2016}, with the help of a
truncated bi-Maxwellian (TBM) model for sheath electron distribution,
that
\begin{align}
q_{en} \approx - \Gamma_{e\parallel}^{se} e\Delta\Phi +
\Gamma_{e\parallel}^{se} \left(T_{e\parallel}^0 - \frac{3}{2}
T_{e\parallel}\right)
\end{align}
where the electron particle flux through the sheath
$\Gamma_{e\parallel}^{se}$ and a nominal temperature
$T_{e\parallel}^0$ are both constants, the potential drop is defined
as $\Delta\Phi(x) = \Phi^w - \Phi(x)$ with $\Phi^w$ the wall
potential, and $T_{e\parallel}(x)$ is the local electron parallel
temperature.  For large ion-electron mass ratio, $m_i/m_e\gg 1,$ the
sheath potential variation is a few times greater than
$T_{e\parallel},$ so the heat flux $q_{en}$ is in the particle flow
direction for $\Delta\Phi <0$ in the sheath. Furthermore, since
$-\Delta\Phi$ drops in magnitude as the wall is approached, the
$q_{en}$ would decrease in magnitude towards the wall as well.
Numerical results in Ref.~\onlinecite{tang_kinetic2016} confirm that
$q_{en} \approx - \Gamma_{e\parallel}^{se} e\Delta\Phi$ is closely
followed in first-principles kinetic simulations, so the spatial
gradient of $q_{en}$ is set by the gradient length scale of $\Phi(x),$
\begin{align}
  \frac{\partial q_{en}}{\partial x} \approx
  -\Gamma_{e\parallel}^{se} e\Delta\Phi \frac{\partial}{\partial x} \ln\left(-\Delta\Phi\right)
  \approx q_{en} \frac{\partial}{\partial x} \ln\left(-\Delta\Phi\right) < 0.
\end{align}
Since the electron density closely follows the Boltzmann distribution in the TBM model, which
agrees well with the first-principles kinetic simulation results,~\cite{tang_kinetic2016} one has
\begin{align}
\frac{\partial}{\partial x} \ln\left(-\Delta\Phi\right) \approx - \frac{T_{e\parallel}^0}{e\Delta\Phi }
\frac{\partial}{\partial x} \ln n_e.
\end{align}
For $- e\Delta\Phi/ T_{e\parallel}^0 \sim 1,$ we have
\begin{align}
 \frac{\partial q_{en}}{\partial x} \sim q_{en} \frac{\partial \ln n_e}{\partial x}. \label{eq-dqn-dx}
\end{align}

\begin{figure}[tbh]
\centering
\includegraphics[width=0.6\textwidth]{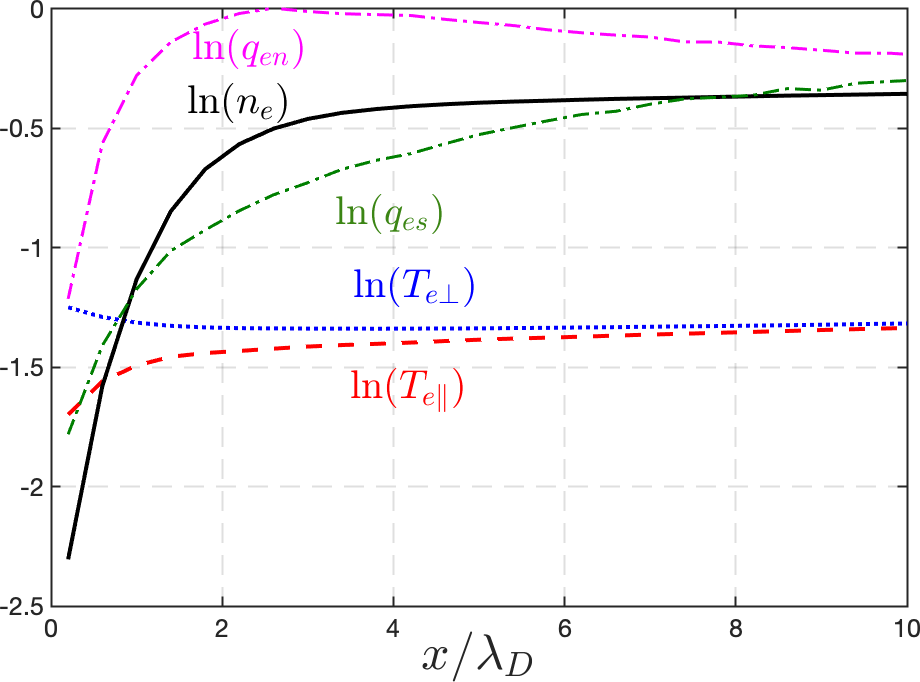}
\caption{Logarithms of $n_e$, $T_{e\parallel,\perp}$ and $q_{en,s}$ near
  the left boundary for the same simulation in
  Fig.~\ref{fig:Teperp-qes-kn-20}. Here $n_e$ is normalized by $n_0$, $T_{e\parallel,\perp}$ by $T_0$, and $q_{en,s}$ by the maximum value of $q_{en,s}$.}
\label{fig:ln-ne-Tepara-Teperp}
\end{figure}

The TBM model~\cite{tang_kinetic2016} completely misses the $q_{es}$
physics as it assumes Maxwellian distribution in $T_{e\perp}$ so it
enforces $q_{es}=0.$ It turns out that $q_{es}\neq 0$ in the cases of
both $Kn^{sh}>1$ and $Kn^{sh}\gg 1,$ but for different physics
considerations.  For a collisional bulk plasma with $Kn^{sh}>1,$ one
can see from the simulations that outside the sheath region, the
collisions are so strong that the total parallel heat flux
$q_x=(q_{en}+2q_{es})/2$ nearly follows the Braginskii's
closure~\cite{Braginskii} $q_x=-\kappa_\parallel dT_e/dx$.  The plasma is also nearly
isotropic outside the sheath as shown in
Fig.~\ref{fig:Teperp-qes-kn-20} where $T_{e\parallel}\approx
T_{e\perp}$. As a result, $q_{es}$ and $q_{en}$ have the same trend as
$q_x$ outside the sheath region, which guarantees that $q_{es}$ has
the same sign with $q_{en}$ even in the sheath region despite the
temperature anisotropy there.
By the heat flux definition of both $q_{en}$ and $q_{es},$ their spatial gradient
is related to the spatial gradient of the distribution function, which can be
written in terms of thermodynamic variables. One insight from the TBM model is that
the dominant terms are
\begin{align}
  \frac{\partial f_e}{\partial x} = n_e \frac{\partial f_e}{\partial
    n_e} \frac{\partial \ln n_e}{\partial x} + T_{e\parallel, \perp}
  \frac{\partial f_e}{\partial T_{e\parallel,\perp}} \frac{\partial
    \ln T_{e\parallel,\perp}}{\partial x}. \label{eq:dfdx-in-n-T}
\end{align}
The plasma potential dependence primarily enters through $n_e,$ but
there is also a contribution through $T_{e\parallel}$ via the parallel
velocity cutoff.  In Fig.~\ref{fig:ln-ne-Tepara-Teperp},
first-principles kinetic simulations confirm the TBM prediction of
\begin{equation}
    \left|\frac{\partial \ln n_e}{\partial x}\right|\gg
    \left|\frac{\partial \ln T_{e\parallel}}{\partial x}\right|,
\end{equation}
and further establishes
\begin{equation}
    \left|\frac{\partial \ln n_e}{\partial x}\right|\gg
    \left|\frac{\partial \ln T_{e\perp}}{\partial x}\right|. \label{eq:density-T-scaling-length}
\end{equation}

Combining Eqs.~(\ref{eq:dfdx-in-n-T}-\ref{eq:density-T-scaling-length}) and from Eq.~(\ref{eq:qes-def}), one finds
\begin{align}
  \frac{\partial q_{es}}{\partial x} \sim \frac{1}{2}\int m_e
  w_\perp^2 w_\parallel n_e\frac{\partial f_e}{\partial
    n_e}d^3\vec{v}\frac{\partial \ln n_e}{\partial x} \sim
  q_{es}\frac{\partial \ln n_e}{\partial x}.\label{eq-dqs-dx}
\end{align}
Recalling the fact that $q_{es}$ has the same sign as $q_{en}$ and $\Gamma_{e\parallel}^{se},$ we come to
the interesting conclusion that the gradient of $q_{es}$ has the same sign
as that of $q_{en}$, which is negative in the sheath region. As a
result, $T_{e\perp}$ will arise towards the wall as predicted by
Eq.~(\ref{eq:Tey-qes-coll}) and confirmed by simulation data in Fig.~\ref{fig:Teperp-qes-kn-20}.

\section{Collisionless plasma with a normal magnetic field to the walls\label{sec-non-coll}}

\begin{figure*}[tbh]
\centering
\includegraphics[width=0.3\textwidth]{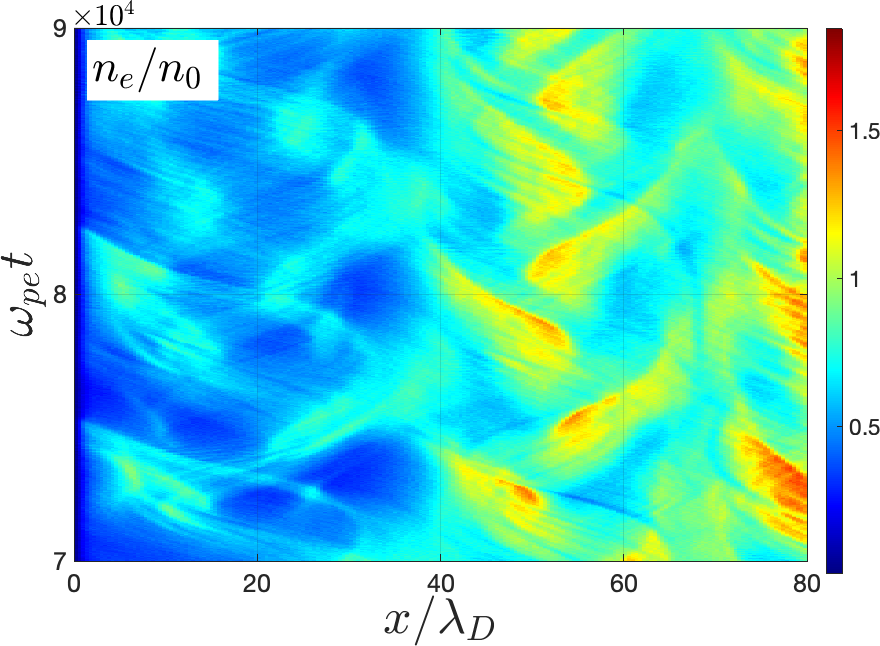}
\includegraphics[width=0.3\textwidth]{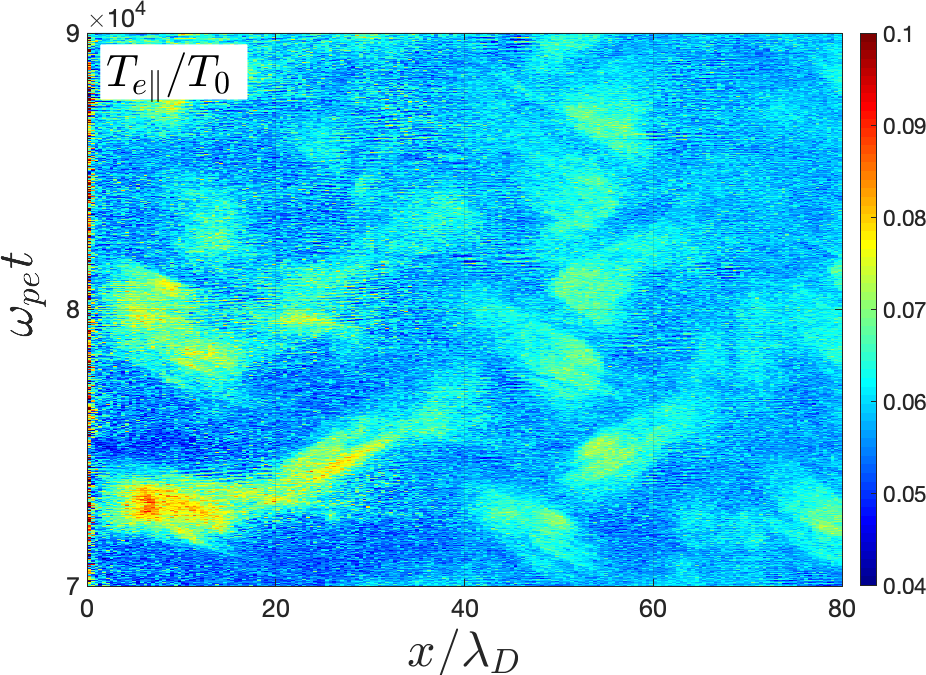}
\includegraphics[width=0.3\textwidth]{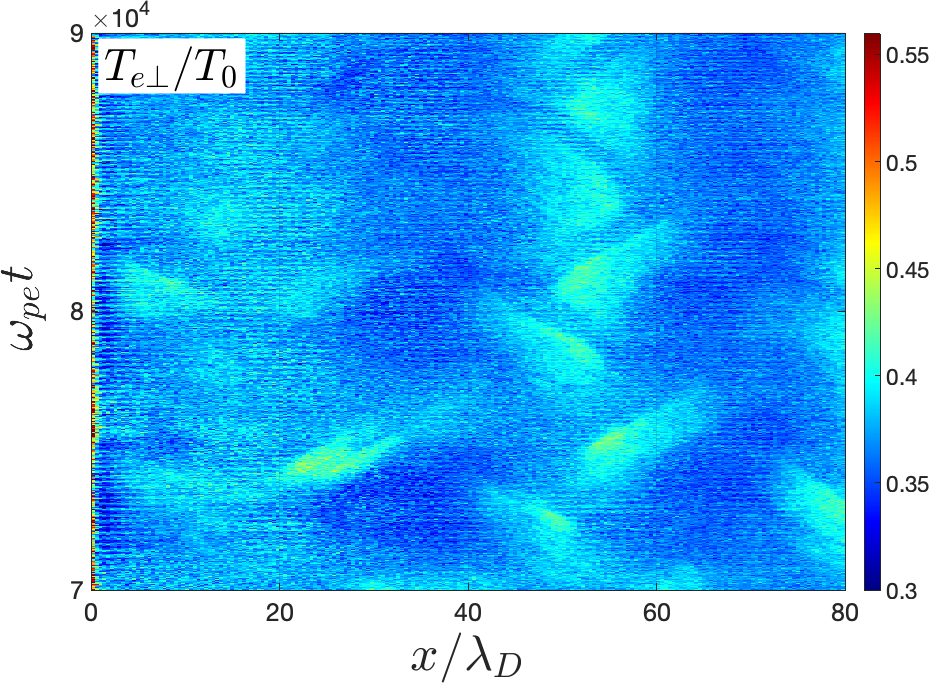}\\
\includegraphics[width=0.4\textwidth]{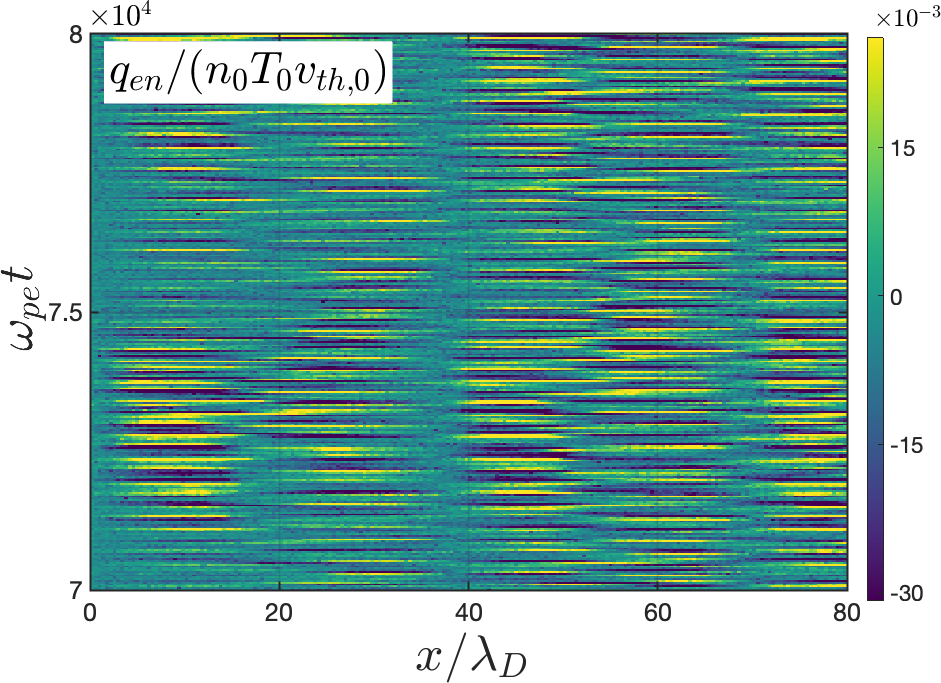}
\includegraphics[width=0.4\textwidth]{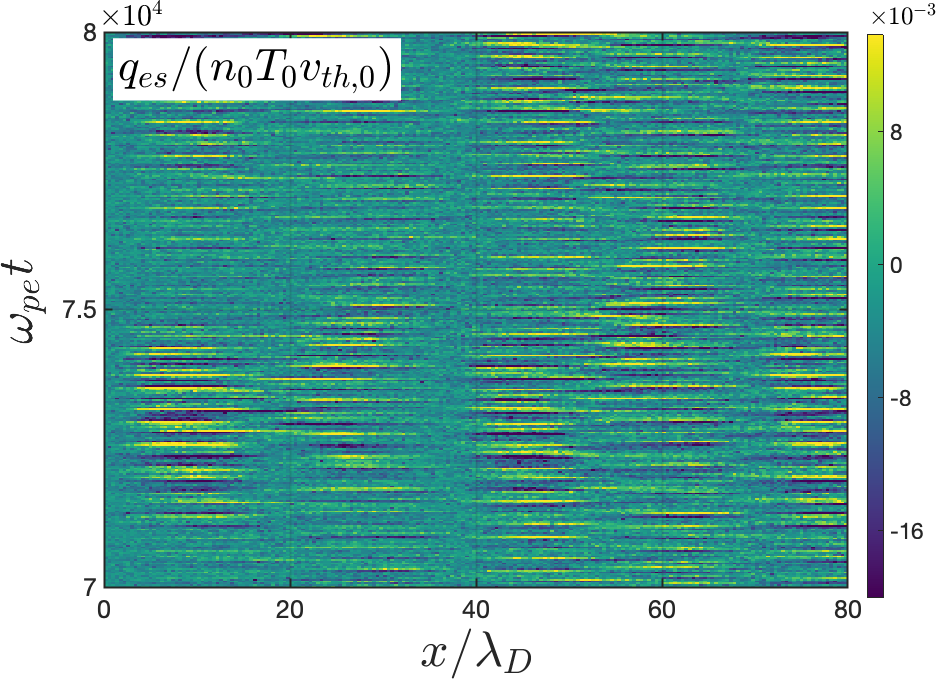}
\caption{Contour plots of electron density, temperature and parallel heat flux
  in a steady state for collisionless plasma in electromagnetic
  simulations with $\beta=1.4\%$. }
\label{fig:Te-collisionless}
\end{figure*}

The limiting case of the $Kn^{sh}\gg 1$ regime is a collisionless
plasma.  In the absence of Coulomb collisions, the pitch-angle
scattering of electrons can be facilitated by the electron interaction
with electromagnetic waves. \textcolor{red}{The most obvious candidate
  of the electromagnetic waves in the sheath problem of $Kn^{sh}\gg 1$
  is the parallel-propagating whistler waves that are robustly excited
  by the electrostatically trapped electrons~\cite{guo2012ambipolar},
  which arise naturally due to the ambipolar potential in the sheath
  region. The Fourier analysis of the perturbed perpendicular magnetic
  field in the VPIC simulation shows that the most unstable mode of
  whistler instability has wavelength $k_x\lambda_D\approx 0.7$ and
  growth rate $\gamma =7\times 10^{-3}\omega_{pe}$ (detailed analyses
  of the dispersion relation and growth rate of whistler instability
  driven by the trapped electrons can be found in Ref.~\onlinecite{guo2012ambipolar}).} Note that
the interaction of electrons with the self-excited whistler waves
causes temperature isotropization by reducing $T_{e\perp}$ from $T_0$,
which is accompanied by a non-zero $q_{es}$. \textcolor{red}{ It is
  worth noting that the role of whistler instability in reducing
  $T_{e\perp}$ and causing its spike in the sheath is further
  highlighted by the comparison of the electromagnetic simulation
  against the electrostatic simulation, where $T_{e\perp}$ remains
  unchanged ($T_{e\perp} =T_0$) and $q_{es} = 0$ in the latter case,
  where $T_{e\parallel}$ is similarly reduced toward the boundary due
  to decompressional cooling.  Such a sharp contrast in $T_{e\perp}$ and $q_{es}$ between the
  electromagnetic and electrostatic simulations also reveals that the
  PIC noise does not induce effective pitch-angle scattering in the
  collisionless limit.}  As in the collisional case, we focus on a
steady state in which the time-averaged plasma state variables remain
nearly constant.

In contrast to the high-$\beta$ plasma (e.g., due to a weak equilibrium
magnetic field as in Ref.~\onlinecite{roberg2018suppression}), the
amplitude of saturated whistler waves is small $\delta B/B_0 \approx
0.1$ in our case for low-$\beta$ fusion plasma with $\beta=1.4\%$. In
our 1D problem, instead of forming whistler turbulence due to the
whistler instability, standing structures, including those of $q_{en,s},$ are
observed in VPIC simulations as shown in
Fig.~\ref{fig:Te-collisionless}, which reinforces the fact that
wave-particle interaction can produce the heat flux $q_{es}.$ However,
in contrast to the electron temperature where the fluctuations are
smaller than the time-averaged values, $\tilde{T}_{e\parallel,\perp}<
\left<T_{e\parallel,\perp}\right>_T$, the electron heat flux is dominated by the fluctuations, $\tilde{q}_{en,s} /
\left<q_{en,s}\right>_T \sim 10$, where the time average
$\left<\right>_T$ is taken in a period much larger than the
whistler period $\Delta T \gg \omega_{ce}^{-1}$. This indicates that
the heat flux is largely independent of the temperature but determined
by the whistler waves~\cite{roberg2018suppression}. It is worth noting
that the interaction of electrons with self-excited whistler waves
cannot completely remove the temperature anisotropy, especially for
low-$\beta$ plasma~\cite{gary1996whistler} as shown in
Fig.~\ref{fig:Te-collisionless}. As a result, $T_{e\parallel}$ is small but $T_{e\perp}$ is
large in the collisionless case compared to those in the collisional
case in Fig.~\ref{fig:Teperp-qes-kn-20}.

\begin{figure}[tbh]
\centering
\includegraphics[width=0.6\textwidth]{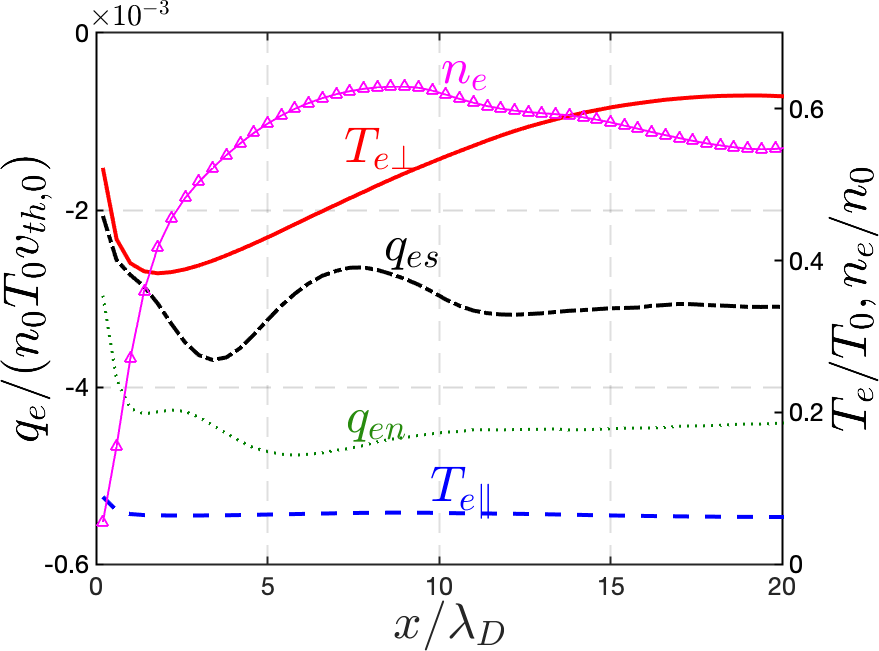}
\caption{Line plots of time-averaged electron temperature
  \textcolor{red}{and density} (right y-axis) and heat flux (left
  y-axis) in a steady state near the left boundary at $x=0$, where the
  period for averaging is $\Delta T = 6.5\times 10^{4}
  \omega_{ce}^{-1}$. They are from the same simulation as shown in
  Fig.~\ref{fig:Te-collisionless}.}
\label{fig:Te-qe-collisionless}
\end{figure}

To get rid of the fast oscillations due to wave-particle interaction,
we can focus on the time-averaged quantities as shown in
Fig.~\ref{fig:Te-qe-collisionless}. For the sake of simplicity, we
ignore the time-averaged symbol $\left<\right>_T$ in the
following. One of the most important findings is that the $T_{e\perp}$
spike near the wall still prevails (although the spike width is slightly smaller), which is still associated with the negative gradient of $q_{es}$. The underlying physics is similar to that in the collisional
case, where the time-averaged energy equations for electron
temperature are similar to Eq.~(\ref{eq:transport-equation-Tex},
\ref{eq:transport-equation-Tey}):
\begin{equation}     
  	n_e
        u_{ex}\ppx{T_{e\parallel}}{x}+2n_eT_{e\parallel}\ppx{u_{ex}}{x}+\ppx{q_{en}}{x}
        = Q, \label{eq:Tex-noncoll}
\end{equation}
\begin{equation}     
  	n_eu_{ex}\ppx{T_{e\perp}}{x}+\ppx{q_{es}}{x}  = - Q/2, \label{eq:Tey-noncoll}
\end{equation}
Here we assumed that the variation of $B$ is so small that $\mathbf{B}$ is still normal to the wall. In addition, the whistler mode is rather coherent in the case considered and thus the fluctuation driven fluxes are tiny compared with the turbulent case, which are ignored compared with their mean values. $Q$
stands for the energy exchange between the perpendicular and parallel
directions due to the wave-particle interaction in which we assume
that there is no energy exchange between the whistler waves and
electrons in the steady state.

Just like the collisional case where the plasma in the
sheath region is nearly collisionless so that $Q_{ee,ei}$ is negligibly
small, $Q$ term should also be ignored in the sheath region due to two
reasons: 1) the deep drop of $T_{e\parallel}$ (via the reduction of
the trap-passing boundary in the electron distribution function) makes
the resonance $\omega-kv_\parallel = \omega_{ce}$ less efficient; 2)
the large spatial gradient of $T_{e\parallel,\perp}$ and $q_{en,s}$ in
the sheath makes $Q$ less important. Then we come to the same striking
realization that \textit{the spike of $T_{e\perp}$ near the wall is
  associated with a negative gradient of $q_{es}$} as in the
collisional case. Considering the similar role of collisions and
wave-particle interaction in the pitch-angle scattering and
temperature isotropization, $q_{es}$ should have similar behaviors as
$q_{en}$ outside the sheath region just like in the collisional case,
which flows from the source region to the walls (e.g., see
Fig.~\ref{fig:Te-qe-collisionless}). So $q_{es}$ would have the same
sign as $q_{en}$ in the sheath. While in the sheath region, the
gradient of the electron density still dominates so that the gradient
of $q_{es}$ has the same sign as that of $q_{en}$, e.g., see
Eqs.~(\ref{eq-dqn-dx}, \ref{eq-dqs-dx}), which is negative. As a
result, $T_{e\perp}$ will spike up near the wall.

\section{The case of an oblique magnetic field intercepting the wall\label{sec-oblique-B}}

There are additional complications in the case of a
collisional plasma in an oblique magnetic field that intercepts
the wall at a small angle. The subtlety can be appreciated by projecting the plasma
energy equations with respect to the magnetic
field~\cite{Chew1956,macmahon1965finite}
\begin{equation}     
  	n_e\mathbf{u}\cdot \nabla T_{e\parallel} +2n_eT_{e\parallel}\nabla_\parallel u_\parallel +\nabla_\parallel q_{en} = 0, \label{eq:transport-equation-Tpara}
\end{equation}
\begin{equation}     
  n_e\mathbf{u}\cdot \nabla T_{e\perp} +n_eT_{e\perp}\nabla_\perp \cdot \mathbf{u}_\perp +\nabla_\parallel q_{es} = 0, \label{eq:transport-equation-Tperp}
\end{equation}
where $\nabla_\parallel =\hat{b}\cdot \nabla$, $\nabla_\perp \cdot
\mathbf{u}_\perp = \nabla \cdot \mathbf{u} - \nabla_\parallel
u_\parallel$.  One can verify this derivation by ignoring the
collisional contribution in the sheath and combining the steady-state
equations for $n_e, P_{e\parallel}\equiv n_e T_{e\parallel},$ and
$P_{e\perp}\equiv n_e T_{e\perp},$ which are given in Eqs.~(3,7,8) of Ref.~\onlinecite{guo-tang-pop-2012a}. A further approximation is that the magnetic field is uniform in the sheath, so $\nabla \cdot \hat{b} =0.$ The electron flow field is allowed to have
components both parallel and perpendicular to the magnetic field,
$\mathbf{u} = u_\parallel \hat{b} + \mathbf{u}_\perp.$ Here we focus on small-$\beta$ plasmas.

If $\mathbf{B}$ is normal to the walls, the electron flow $\mathbf{u}$
is aligned with the magnetic field $\mathbf{B}$ so that
$\mathbf{u}_\perp =0$. As a result, we recover
Eqs.~(\ref{eq:transport-equation-Tex},
\ref{eq:transport-equation-Tey}). Whereas, for an oblique magnetic
field with a small angle to the walls, the sheath field acceleration
is bending the ion flow in the Chodura layer from the direction
aligned with the magnetic field towards the direction normal to the
wall. In the analogous electron Chodura layer, which is next to the
wall and of electron gyroradius in width, the electron flow
$\mathbf{u}$ also tilts away from the magnetic field line and toward
the wall, so that $\nabla_\perp\cdot \mathbf{u}_\perp \neq
0$. This non-zero divergence of the perpendicular
electron flow, as we will show below, provides an additional cooling
mechanism that can overcome the heating effect of $\nabla_\parallel
q_{es}.$

\begin{figure*}[htb!]
\centering
\includegraphics[width=0.45\textwidth]{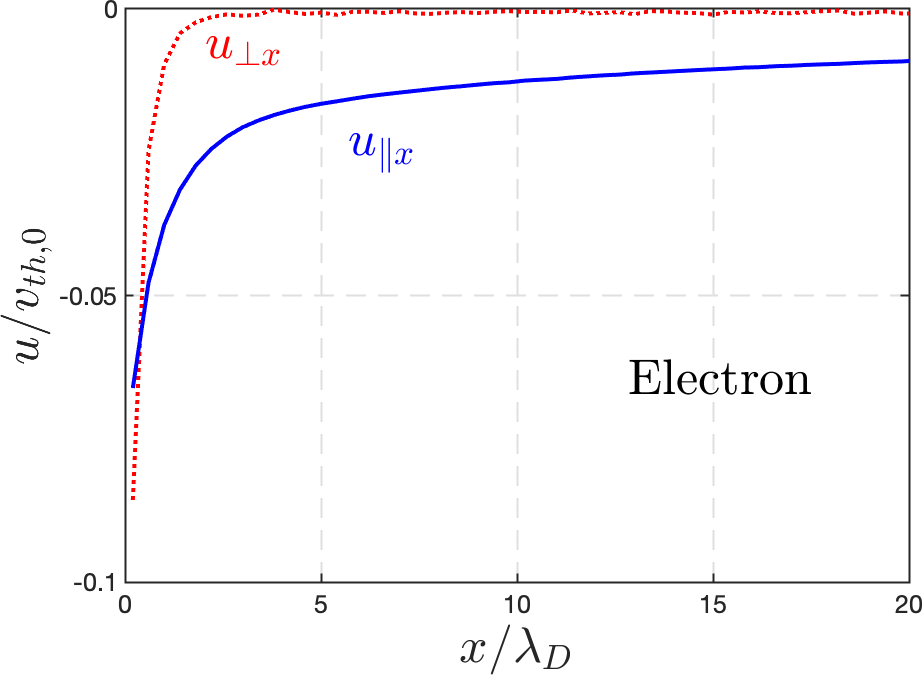}
\includegraphics[width=0.45\textwidth]{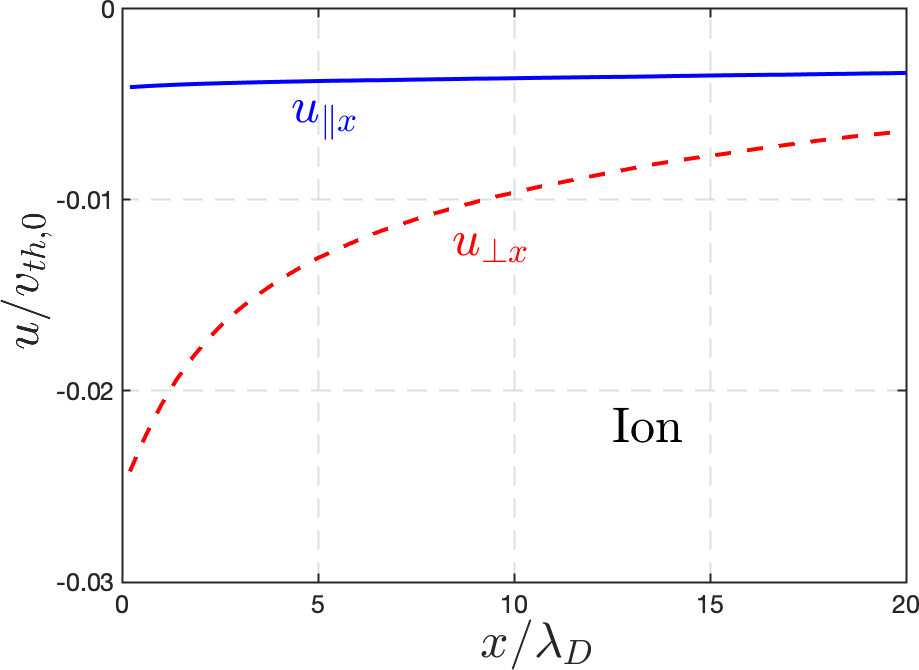}
\caption{The projections of the parallel and perpendicular plasma flow onto the x-direction. The simulation setup is the same as Fig.~\ref{fig:Teperp-qes-kn-20} but with an oblique magnetic field with $\theta=10^\circ$. }
\label{fig:vx-angle-10}
\end{figure*} 

\begin{figure}[tbh]
\centering
\includegraphics[width=0.6\textwidth]{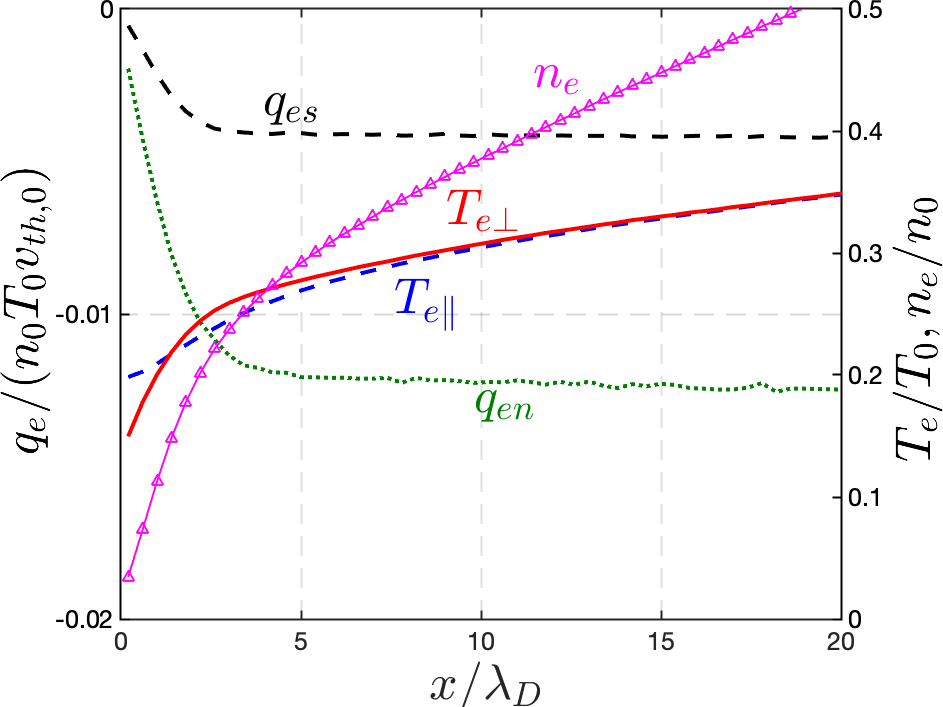}
\caption{The electron temperature \textcolor{red}{and density} (right y-axis)
  and heat flux (left y-axis) in a steady state
  for the simulation in Fig.~\ref{fig:vx-angle-10}.}
\label{fig:Teperp-qes-kn-20-tilde_B}
\end{figure}

Let the wall-intercepting magnetic field $\mathbf{B}$ lie in the x-y
plane, and take the form $\mathbf{B}= B \sin(\theta) \mathbf{e}_x +
B\cos(\theta)\mathbf{e}_y$ with a small angle $\theta$, where the wall is
in the y-z plane. Assuming that the plasma is uniform in the y-z
plane, all the spatial derivatives can be projected into the x-direction:
$\mathbf{u}\cdot \nabla = u_x \nabla_x$, $\nabla_\parallel
=\sin(\theta) \nabla_x$, and $\nabla_\perp =\mathbf{e}_x\cos(\theta) \nabla_x$. As a result, $\nabla_\parallel
u_\parallel = \nabla_x u_{\parallel x}$ and $\nabla_\perp \cdot
 \mathbf{u}_\perp = \nabla_xu_{\perp x}$ with $u_{\parallel, \perp x}$ being the
projection of the parallel plasma flow and perpendicular flow in the x-y
plane onto the x-direction. Notice that for $\theta=\pi/2$ we readily
recover the normal magnetic field case. For an oblique magnetic field
with a small angle $\theta$, $u_{\perp x}$ will increase approaching the
wall due to the electron pressure drive (e.g., see
Fig.~\ref{fig:vx-angle-10}) so that $\nabla_\perp \cdot
\mathbf{u}_\perp >0$ in Eq.~(\ref{eq:transport-equation-Tperp}), which
provides decompressional cooling. In contrast to ions, where the
acceleration of the projected flow in the sheath is mainly via
$\mathbf{u}_\perp$ due to the sheath electric field, both the parallel and
perpendicular electron flows contribute nearly equally to the plasma
flow at x-direction as shown in Fig.~\ref{fig:vx-angle-10} since their
acceleration drive is the electron pressure. As a result, $\nabla_x
u_{\perp x} \approx \nabla_x u_{\parallel x}$ for electrons so that the decompressional cooling for $T_{e \parallel}$ and $T_{e\perp}$ would be similar, which overwhelms the heating due to $\nabla_\parallel q_{en,s}$. Therefore, there is no spike in the parallel and
perpendicular electron temperature and thus any other projected
temperature (e.g., see Fig.~\ref{fig:Teperp-qes-kn-20-tilde_B}).

\section{Conclusion \label{sec-conclude}}

In conclusion, the long-standing mystery of $T_{e\perp}$ spike has been resolved in the
non-neutral sheath region in an unmagnetized plasma or a magnetized plasma with a nearly normal magnetic field to the walls, which is found to be associated with a negative gradient of the electron heat flux $q_{es}$. Such a non-zero heat flux is induced by either the
interaction of electrons with the self-excited whistler waves in a
nearly collisionless plasma or the collisions (or both of them in the
intermediate regime), both of which result in temperature
isotropization by reducing $T_{e\perp}$ from the initial value
$T_0$. The former has an electromagnetic nature, while the latter
works equally for the electrostatic and electromagnetic models of
plasmas. It has been shown that the negative gradient of $q_{es}$ is
related to that of $q_{en}$ in the sheath region due to the strong
drive of $\partial n_e/\partial x$ as a result of the large ambipolar
potential. However, for an oblique magnetic field intercepting the wall with a small angle, the situation is different in that the decompressional cooling will overwhelm the heating due to the electron heat flux in both the parallel and perpendicular directions. As a result, there will be no spike in the electron temperature in any direction.

We thank the U.S. Department of Energy Office of Fusion Energy
Sciences and Office of Advanced Scientific Computing Research for
support under the Tokamak Disruption Simulation (TDS) Scientific
Discovery through Advanced Computing (SciDAC) project at both Virginia
Tech under grant number DE-SC0018276 and Los Alamos National
Laboratory (LANL) under contract No.~89233218CNA000001. Y.Z. was
supported under a Director's Postdoctoral Fellowship at LANL. This
research used resources of the National Energy Research Scientific
Computing Center (NERSC), a U.S. Department of Energy Office of
Science User Facility operated under Contract No.DE-AC02-05CH11231.

\bibliography{reference}


\end{document}